\documentclass[english]{article}
\usepackage[T1]{fontenc}
\usepackage[latin9]{inputenc}
\usepackage{textcomp}
\usepackage{footnote}
\usepackage{amsmath}
\usepackage{amssymb}
\usepackage{stmaryrd}

\makeatletter

\makesavenoteenv{description}

\@ifundefined{date}{}{\date{}}
\makeatother

\usepackage{babel}
\begin{document}
\title{Logic, Language, and Calculus\\
{\normalsize{}Logical Understanding I}}
\author{Florian Richter}
\maketitle
\begin{abstract}
The difference between object-language and metalanguage is crucial
for logical analysis, but has yet not been examined for the field
of computer science. In this paper the difference is examined with
regard to inferential relations. It is argued that inferential relations
in a metalanguage (like a calculus for propositional logic) cannot
represent conceptual relations of natural language. Inferential relations
govern our concept use and understanding. Several approaches in the
field of Natural Language Understanding (NLU) and Natural Language
Inference (NLI) take this insight in account, but do not consider,
how an inference can be assessed as a good inference. I present a
logical analysis that can assesss the normative dimension of inferences,
which is a crucial part of logical understanding and goes beyond formal
understanding of metalanguages. 
\end{abstract}

\section{Introduction}

\subsection{Context -- Commitments}

Language is not only descriptive, it is also normative. Just by describing
language and how it is used, i.e. creating models from language use,
it will not be possible to recreate an ability that yields understanding.
That language is normative does not only mean that language use is
guided by norms, but it also implies a certain assessment of what
is correct and what is incorrect language use. Dictionaries and grammar
books encapsulate the correct use of a natural language, so does logic
for artificial as well as for natural languages.

\subsection{Problem}

Approaches in the field of natural language understanding focus on
the descriptive part of language use. They describe how language is
used. It can lead to a certain understanding or better representation
of the language usage of a group and depict maybe also to a certain
degree a semantic component, but it will not yield a more complete
understanding of semantic significance, unless also the normative
component is considered. Normativity is a key part of meaning. 

We play different language games that are governed by different norms.
Our assessment of the normative significance guides our understanding.
Normative assessment of meaning is something exceptionally done by
humans, but can it be also computed or to a certain degree expressed
in mathematical models? First and foremost it has to be examined,
how it can be expressed logically in order to see how it can be implemented
in a mathematical model.

\subsection{Motivation -- The \emph{Status} of Logic}

It is important to distinguish between the \emph{status} of the forms
of languages that are used. The distinction between object-language
and metalanguage is crucial, because the metalanguage allows the possibility
to talk about the object-language and the concepts that cannot be
expressed in the object-language. Paul Hoyningen-Huene calls the metalanguage
of statement logic (object-language) ``metalogic of statement logic''.\cite{Hoyningen-Huene2004}
It allows one to express the concepts of ``logical truth'' and ``valid
inference'' for statement logic. Both concepts are determined by
the use of the operators in statement logic and are purely syntactical.
The logical truths of statement logic are therefore also only tautologies.
It is possible to build a calculus of statement logic as a metalanguage.

Scientific examinations are written in natural language, but it is
not directly a metalanguage in the logical sense, because it describes
examined objects, although e.g., you can write an examination about
German grammar in English. English is then your metalanguage that
you use to describe German: the object-language. The difference between
metalanguange and object-language can also be used to clarify, about
what one talks, i.e. what is the object: an object/entity or an expression
(e.g. the name of a city or the city itself). \emph{But} when it comes
to describe and assess the logic of language, this distinction faces
several problems. 

Logic is not just a formal tool that formulates a calculus of an object-language
in an artificial metalanguage, it is part of the use of language.
The attempt of this paper is to understand the \emph{status} of logic
in language use. I believe that making explicit this status will not
only lead to a deeper understanding of logic and language, but will
also enable us to build mathematical models of natural language that
can be used in machine learning.

\section{Philosophical Background}

Kurt Gödel writes that ``every precisely formulated yes-and-no question
in mathematics must have a clear-cut answer.''\cite{Godel1995} (379)
Gödel was a Platonist and believed that there is a universe of all
possible discourse, when it comes to mathematical objects. This ontological
part has to be considered. Many problems about the foundations of
mathematics in the first half of the 20th century were centered around
the connection of set theory and logic as the fundament of mathematics.\cite{Bedurftig2012}
Logic gives in a certain way the syntactical or formal structure and
set theory the semantical or ontological component.\footnote{The disjunction ``or'' should not mean that these terms can be used
as synonyms. This connection is much more complicated.} It depends in a way then on one\textasciiacute s ``ontological commitment''
to mathematical objects and how big the ontological realm should be,
i.e. what it all includes, e.g. physical objects, mathematical objects,
and so forth. Willard Van Orman Quine sees them as ``myths'' that
are ``good and useful'', because they simplify our theories.\cite{Quine1953}
The problem is that these mathematical objects should serve as truth-makers
of mathematical propositions and if there is a possible universe of
all mathematical objects, then this implies that the propositions
are decidable. It seems like Gödel had the idea that such a decidability
for mathematics could be possible\cite{Bedurftig2012}, despite his
incompleteness theorems. -- It is not the case that there exists
a universe of all possible discourse with truth-makers for propositions
or concepts. We do not discover the meaning of them and there can
also not be some kind of ``metaphysical glue''\cite{Putnam1992}
(109) that attaches a word to a referent or object that conveys or
gives meaning to the word or also to a proposition.

Gödel\textasciiacute s incompleteness theorems stem from the phenomenon
of self-reference, i.e. a system is powerful enough to talk about
itself or at least to name objects. The system or mathematical model
is able to formalize mathematical objects via natural numbers (the
so called ``Gödel numbering''). Every arithmetic formula is coded
as a number and is assigned a definite Gödel number. The Gödel numbering
is now used to formulate formulas that are expressively powerful enough
to define concepts like ``provable formula'' and then also the negation
of this concept.\cite{Godel1931} (175/176) It is sometimes expressed
in the following way:
\begin{description}
\item [{F}] = ``I am not provable.''\cite{Ertel2017}
\end{description}
Gödel writes that we have a sentence that asserts its own unprovability.\cite{Godel1931}
(175) ``Suppose, F is wrong. Then F can be proven and has therefore
shown that F is not provable. This is a contradiction. Thus, F is
true and therefore not provable.''\cite{Ertel2017}

Of course, it is a mix up of the predicates ``true'' and ``provable''.
Gödel plays with the similarity of the meaning ``provable'' and
``true''. The former is a property of formulas and the latter a
property of sentences. Gödel also mentions that these ideas are cognate
with the liar paradox, where it is just about the truth of statements.\cite{Godel1931}
Tarski developed this problem further in the field of semantics towards
a formalized theory of truth.\cite{Tarski1936} Both contributions
to logic lie on the possibility of building a metalanguage to speak
about an object-language.

It is problematic to hang the reflexivity of language on the difference
between metalanguage and object-language, because the metalanguage
consists of (abstract) names (metavariables) for the object-language,
where it is about the use of language. One of the consequences of
Gödel\textasciiacute s incompleteness theorems is that the construction
of a metalanguage with \emph{well-defined names} for predicates like
``provable'' or ``true'' is not possible, because it leads to
contradictions. It has not been possible to develop a well-defined
metalogical or metamathematical model with well-defined names for
``provable'' or ``true'' and despite that the mathematicians use
proofs and despite that the philosophers use the predicate ``true'',
so it does not seem to confound us in our use of these words or better
in the language games that we play. -- I believe that the reflexivity
of language stems from an epistemic gap between holding something
for true and that something is true (or between seeming right and
being right).\cite{Wittgenstein2006} (§§ 293 and 303) The possibility
to err or to be wrong opens up the language game that we play. (How
this is connected with modal logic, will be explained below.)\\
\\
Artificial languages and formalization help to disambiguate the meaning
of sentences and to understand the important distinction between metalanguage
and object-language. This logical distinction can be sometimes seen
as pedantic, but it is important to not get lost in the debates in
linguistics, computer science, mathematics, and philosophy. It has
to be clear, if it is just a name or if the word is used in a language
and also how it is used. Naming something is also using a word (like
e.g. predicating). Peter Geach pointed out that there is a difference
between calling a thing ``P'' and predicating of a thing ``P''.
It may be that the predication ``P'' of a thing is embedded in an
if-then-clause or in a disjunctive proposition\cite{Geach1960} and
then it is embedded in a logical context.\footnote{Another example that reflects the ideas of this paper is: If there
is \emph{not} a universe of all possible discourse with meaningful
facts, that warrants objectivity, then meaning does not have to be
subjective.} 
\begin{quote}
``To say, \textquoteleft If the policeman\textasciiacute s statement
is true, the motorist touched 6o mph' is not to call the policeman\textasciiacute s
statement true; to say, \textquoteleft If gambling is bad, inviting
people to gamble is bad' is not to call either gambling or invitations
to gamble \textquoteleft bad.'{}''\cite{Geach1960} (223)
\end{quote}
True is used here as a descriptive predicate. It makes the content
judgeable and does not solely take it as a speech act (ascription).\cite{Brandom2009}
(31/32)\footnote{Brandom emphasizes this point specifically for the connection of philosophy
of language, cognitive science, and artificial intelligence in the
chapter ``How Analytic Philosophy Has Failed Cognitive Science''.} The if-then-clause specifies also under which circumstances it should
be correct to call the policeman\textasciiacute s statement true.
It gives to the descriptive content a normative assessment, because
``true'' is a normative concept and not a descriptive concept. If
one commits oneself to the antecedent, one also has to commit onself
to the consequent. Committing onself is a normative doing along with
being entitled to a claim.\cite{Brandom2008}\footnote{Commitment and entitlement are deontic concepts. See \cite{Brandom1994}} 

If we want to understand what meaning means, we cannot just look at
the manipulation of symbols. This would be just manipulating strings
of \emph{names} connected by logical operators. At least this is the
valid insight of John Searle\textasciiacute s thought experiment about
the chinese room. The outcome of the argument is that understanding
natural languages is something more than to follow rules or instructions.\cite{Searle1980}
\emph{But then how can the trick be done?} This question is not answered
by Searle. He believes that we first need to understand the brain
to build something like a strong artificial intelligence that understands
natural languages. It might be also a valid way to first look at technological
advancements, to understand how the trick is done. Therefore, one
should look into the field of ``Natural Language Understanding''
(NLU) in computer science.

\section{Literature Review}

\subsection{General Outlook -- Natural Language Understanding}

There are different forms of understanding that have to be considered.
There are approaches that group together expressions that have similar
meanings. This can be done by vector space models of semantics. (For
a general overview of these models see Turney et al. (2010).\cite{Turney2010})
Another form of understanding is to be able to answer queries about
a text. Hermann et. al. (2015) present a model that can identify objects
(expressions) within a text as answers to questions.\cite{Hermann2015}
Evans et al. (2018) propose a model for ``recognizing entailment
between logical formulas''. Of course, they also state that there
is a difference between recognizing entailment between logical formulas
and ``recognizing entailment between natural language sentences''.
``Evaluating an entailment between natural language sentences requires
understanding the meaning of the non-logical terms in the sentence.''
They seem to have the idea that first the formal logical understanding
of the model has to work, before one can apply it in natural language.\footnote{``We believe that isolating the purely structural sub-problem will
be useful because only networks that can reliably predict entailment
in a purely formal setting, such as propositional (or first-order)
logic, will be capable of getting these sorts of examples consistently
correct.''}\cite{Evans2018} (10) A crucial point is, how do they now that the
entailments are correct or valid? The write that ``{[}e{]}ntailment
is primarily a semantic notion: \emph{A} entails \emph{B} if every
model in which \emph{A} is true is also a model in which \emph{B}
is true.'' This is a definition in the metalanguage of the calculus
of propositional logic and in order to ``test if $A\models B$,''
they ``test whether $A\land\lnot B$ is satisfiable''.\cite{Evans2018}
(2/3) In the metalanguage of the calculus it is tested, whether there
is an inconsistency or not. The normative assessment is solely based
on the principle of avoiding inconsistency and establishing a consistent
calculus. 

\subsection{Natural Language Inference}

In 2006 Dagan et. al wrote one of the first papers in the field of
Natural Language Inference (NLI). They stated ``that textual entailment
recognition is a suitable generic task for evaluating and comparing
applied semantic inference models.'' And also hoped that ``{[}e{]}ventually,
such efforts can promote the development of entailment recognition
\textquoteleft engines' which may provide useful generic modules across
applications.''\cite{Dagan2006}

Bowman et al. (2015a) elaborate a neural network model of a ``\emph{relational}
conception of semantics'' as a counterpart to distributed semantic
representation. The meaning is governed by the inferential conncetions:
\begin{quote}
``For instance, \emph{turtle} is analyzed, not primarily by its extension
in the world, but rather by its lexical network: it entails \emph{reptile},
excludes \emph{chair}, is entailed by \emph{sea turtle}, and so forth.
With generalized notions of entailment and contradiction, these relationships
can be defined for all lexical categories as well as complex phrases,
sentences, and even texts. The resulting theories of meaning offer
valuable new analytic tools for tasks involving database inference,
relation extraction, and textual entailment.''\cite{Bowman2015a}
\end{quote}
The Stanford Natural Language Inference (SNLI) corpus\cite{Bowman2015}
contains 570k sentence pairs with the labels entailment, contradiction,
and neutral and is used e.g. also by Rocktäschel et al. (2015)\cite{Rocktaschel2015}
and Cases et al. (2017)\cite{Cases2017}. Nie et al. (2019) focus
on another aspect of NLI: the elaboration of a data collection method.\cite{Nie2019}
Geiger et al. (2018) propose a ``method for generating artificial
data sets in which the semantic complexity of individual examples
can be precisely characterized''. The ``method is built around an
interpreted formal grammar that generates sentences containing multiple
quantifiers, modifiers, and negations''.\cite{Geiger2018} For an
approach that relies on artificial data sets see also Geiger et al.
(2019)\cite{Geiger2019}. There have been also studies about monotonic
inferences through ``interactions of entailment reasoning with negation''.\footnote{``We would like to determine whether a system can actually reason
about lexical entailment and, furthermore, whether it has learned
that negation is \emph{downward monotone} (roughly, that A entails
B if, and only if, \emph{not}-B entails \emph{not}-A, for all A and
B).''\cite{Geiger2020} (2)} Geiger et. al. (2020) call it ``Monotonicity NLI''.\cite{Geiger2020} 

\section{Problems of Reasoning}

The SNLI corpus from Bowman et al. (2015) contains statements that
are based on descriptions of images. Entailment, contradictory, and
neutral statements were compiled and verified, but it is hard to reason
with inferential relations that were made like the ones in the corpus,
because logical connectors govern or should govern the correctness
of inferences. The database is a descriptive representation of language
and does not include a normative assessment of the inferential relations
per se. It is something different to follow the rule and to be able
to assess the correctness of the rule. That is why Ludwig Wittgenstein
says that rule-following is a practice, because the mastery of a practice
is something else than just blindly following a rule. 

If one claims that e.g. x is a turtle, then it might make sense to
claim that this claim excludes the claim that x is a chair (a contradictory
statement). And it makes sense to infer that, if x is a turtle, it
is an animal (entailment statement). One could now ``infer'' that
a chair is also not an animal. That is correct, but if the contradictory
statement would be the claim that x is a bird, one could make the
inference that x is not an animal, which is not correct. 

That a turtle is an animal is a form of inductive reasoning. It cannot
be just inferred whether a bird is an animal or not. It relies on
other inferences and facts -- or on a semantic web that yet has not
been established, but should be learned. In deductive reasoning semantic
relations can be established more easily. If animal is contradictory
to furniture, then all that is incompatible with animal will also
be incompatible with the more specific concepts (like e.g. bird and
turtle) that fall under the more general concept. But for deductive
reasoning also a semantic web has to be already established.

Robert Brandom introduces the idea of an incompatibility semantics
that is more precise with the logical vocabulary.\cite{Brandom2008}
(121/122) Contradiction in formal logic has a different sense than
the use of contradiction in natural language inferences in models
of machine learning. Contradiction and entailment have already a (logical)
meaning that governs their use. One needs to be precise of the use
of concepts on a logical level otherwise the use of non-logical vocabulary
is even more difficult. Of course, in natural language we often understand
the ambiguous use of words or even the strange use of non-logical
vocabulary, because we are mastering the practice of speaking and
understanding the language and we can (at least most of the time)
make sense of it. -- \emph{I believe that the logic, that governs
inferences, can be made explicit by modal vocabulary, which hopefully
can be represented in mathematical models.}\footnote{Brandom shows, how normative and deontic vocabulary (commitment and
entitlement) can make modal vocabulary explicit. Participating in
practices of giving and asking for reasons (as Wilfrid Sellars puts
it) for the language games that we play. We are being held responsible
for our claims as commitments and have to justify, why and how we
are entitled to these commitments.\cite{Brandom1994,Brandom2008}
This is a crucial part of a pragmatic approach to language and human
understanding, but it goes beyond what can be represented in mathematical
models.}

\subsection{Modal Logic}

Modal logic is an extension of propositional logic. The first extension
is the so called system \textbf{K} (after Saul Kripke) and the next
extension is the system \textbf{T}. This system introduces the following
axiom:
\begin{description}
\item [{$\boxempty\varphi\rightarrow\varphi$\footnote{I use greek letters for formulas, which belong to the metalanguage
of propositional calculus. Latin letters are for statements of the
object-language.}}]~
\end{description}
With this axiom the following theorem can be proven:
\begin{description}
\item [{$\varphi\rightarrow\diamondsuit\varphi$}]~
\end{description}
Statements or compounded/combined statements are possible, if they
are statements. This marks an important difference to propositional
logic, because to take a statement as a possible statement, is taking
it as a move within a language game of giving and asking for reasons
by making explicit e.g. entailments of the statement that can serve
as reasons. This is different to the usual semantic approaches to
modal logic that rests on the notion of possible worlds. 

Kripke introduces the idea of possible worlds to represent the semantics
of modal logic. It is possible that a statement is true, if there
exists at least one possible world in which the statement is the case.
The possible worlds can stand in relations, which means that they
are epistemically accessible.\cite{Kripke1963} We could take e.g.
our world as the initial possible world and think of counterfactual
situations that represent other epistemically accessible possible
worlds. The system \textbf{T} has an important characteristic, if
it is understood within the framework of a possible world semantics.
The system \textbf{T} is reflexive, i.e. that the possible world is
accessible to ``itself'' (wRw) or as Kripke writes: ``It is clear
that every world H is possible relative to itself; for this simply
says that every proposition true in H is also possible in H.''\cite{Kripke1963}
(70)

An analogy of a card game might be helpful to explain the idea of
possible worlds and the epistemic accesibility.\cite{Zoglauer2002}
(132) A player knows her own cards, but does not know the cards of
all the other players, although it might be, that the player knows
some cards of other players (epistemically accesible worlds) and it
is possible for the player to imagine some cards of the other players
(counterfactual possible worlds). If she knows her own cards, then
this could also count as a certain kind of consciousness of her own
(epistemic) states. This kind of reflexivity is different to the difference
of metalanguage and object-language, that was mentioned above, which
only originates from the possibility of naming things in a metalanguage. 

For statement or propositional logic a calculus can be developed by
purely syntactical means.\cite{Hoyningen-Huene2004} We have therefore
a metalanguage that is purely syntactical. The metalanguage contains
words like proofability and derivability, but these concepts have
clearly also a modal background. Rudolf Carnap tried to construct
a syntactical metalanguage for modal logic\cite{Carnap1937} (250/151),
while e.g. Quine argued that modal logic opens an intensional context
that is opaque.\cite{Quine1980} Therefore, a calculus in a syntactical
metalanguage cannot be developed. I will here not argue for one or
the other side, but propose a different approach.

It is an important step in understanding that statements are only
possibly true and if they are possibly true that there is at least
on possible world that is epistemically accessible. Language is a
social practice and different players interact to know which statement
is true. Introducing modal vocabulary is a step in realizing that
one is a player in that game with a certain epistemic states (possible
worlds). The epistemic states stand in possible relations with each
other. This relations can be expressed as incompatibility and compatibility
relations.

\subsection{Incompatibility}

The modal vocabulary operates in the scope of nonmontonic inferences,
like e.g. material inferences. They express good inferences based
on the principle of material incompatibility. The claim, if p then
q, is incompatible with the claim that it is \emph{possible} that
p and not-q.\cite{Brandom2008}
\begin{description}
\item [{$p\rightarrow q$}] is incompatible with $\diamondsuit(p\land\lnot q)$
\item [{$p\rightarrow q$}] is compatible with $\diamondsuit(\lnot p\lor q)$
\item [{$p\leftarrow q$}] is compatible (equivalent) with $\diamondsuit(p\lor\lnot q)$
\end{description}
Another logical equivalence is the law of contraposition $(p\rightarrow q)=(\lnot q\rightarrow\lnot p)$.
This law expresses a certain similarity to one tautology in statement
logic: the \emph{modus tollens} ($((p\rightarrow q)\land\lnot q)\rightarrow\lnot p$).
So, if you know that not-q is the case, you can infer that not-p is
the case, which expresses incompatibility relations of entailments
that go from a specific claim to a more general claim, e.g. if x is
a turtle, then it is an animal. And what is not an animal is also
not a turtle. 

With these material incompatibilities it can be seen, why chair is
a better contradictory statement to turtle than bird, because it supports
more good inferential relations. It lies also on a similar conceptual
level. Chair entails furniture and everything that is incompatible
with furniture is a also incompatible with chair and what is incompatible
with animal is incompatible with turtle. The only problem is that
a chair has (often) four legs and so does a turtle. It is a common
property (of course there are different kinds of chairs, but let us
leave that aside, because we cannot assume that a algorithm ``knows''
something like that). So, it cannot be that everything is incompatible
in this case. 

\subsection{Entailment}

What does now entailment mean? It has to be distinguished between
different kinds of entailments -- there is e.g. the \emph{modus ponens}
(metalanguage) or the implication (object-language) -- to clarify
the concept of entailment. The \emph{modus ponens} is a deductive
form of reasoning: 
\begin{description}
\item [{(1)}] $\varphi$ and $\varphi\Rightarrow\psi$ therefore $\psi$ 
\end{description}
The \emph{modus ponens} is sometimes called the ``implication elimination'',
because it allows to eliminate the implication and to detach the consequent
of the implication, i.e. to assert it. With regard to Gerhard Gentzen
and his calculus of natural deduction it can be represented the following
way\cite{Gentzen1935} (186):
\begin{description}
\item [{(2)}] $\frac{\varphi,\varphi\Rightarrow\psi}{\therefore\psi}$
\end{description}
Deductive reasoning is certainly a sound way of reasoning, but the
\emph{status} of it can nevertheless be questioned. Lewis Carroll\textasciiacute s
story of the tortoise and Achilles gives an insight about the difference
of metalanguage and object-language and the specific status that the
\emph{modus ponens }has. The tortoise does not accept the consequent
in a deductive inference and is not convinced by the application of
the rule of inference to detach the consequent. The tortoise adds
another premise that represents the rule of inference in the object-language
(``If A and B are true, Z must be true.''), but does still not accept
the consequent and goes on to add more premises that supposedly should
represent the rule of inference to finally detach the consequent.\cite{Carroll1895}
Betrand Russell discusses these ideas by distinguishing between the
assertion of propositions and the meaning of ``\emph{therefore}''
in the metalanguage on the one side and the meaning of ``\emph{implies}''
in the object-language on the other side.\cite{Russell1996} (§ 38)
According to Gilbert Ryle there is a difference between applying a
rule of inference, which the tortoise does not do (a kind of knowing
how) or refuses to do and the acknowledgement of the propositional
content (a kind of knowing that).\cite{Ryle} (It seems that, for
Ryle, to know how to do something or to apply it is prior to the knowledge
of the rules or the propositional content.) The difference between
metalanguage and object-language has to be considered to understand
the \emph{status} of the inferences. I believe, it is therefore wrong
to directly go from the \emph{modus ponens} to implications like e.g.
Friedrich Kambartel and Pirmin Stekeler-Weithofer suggest\cite{Kambartel2005}
(212/213):
\begin{description}
\item [{(3)}] $p$ and $p\rightarrow q$ then $q$
\end{description}
The first formulation (1) is a rule of inference in the metalanguage,
while the other formulation (3) is within the object-language of propositional
logic. Even the attempt to write it similar like the rule of inference
does fall short of the difference in scope and use, because the sentence
or proposition $q$ cannot be detached and the implication cannot
be eliminated in the object-language. The correct formulation would
be:
\begin{description}
\item [{(4)}] $((p\rightarrow q)\land p)\rightarrow q$
\end{description}
It is a compounded sentence in the object-language that cannot be
dissolved in seperate propositions. It is one proposition. The usage
of the rule of inference in a meta-language, like e.g. the calculus
of propositional logic, allows to use the detached and derived consequent
as a premise and add it to the ``true'' or valid ``set'' of propositions
($\varSigma$). The requirement for propositional logic is then that
a contradicton cannot be derived, because this would lead to an inconsistent
or incoherent ``set'' of propositions. At least for propositional
logic such a calculus can be developed.

A sentence only has meaning, if it is embedded in a logical context
by the logical connector. The inferential relations show which role
the sentences plays, if it is a premise or a consequent. In a metalanguage
the derived consequent could also serve as a premise, but that would
be to abstract from the original role it played and to not consider
the context or circumstances under which it was implied. The propositions
in a metalanguage are just names of propositions (hence the greek
letters to make that difference). They are not used and have no meaning.
They are abstract and empty placeholders. Only in the logical context
of an object-language they can have meaning. Propositions cannot be
detached from their conditions, otherwise it would also not be an
inferential semantics. In a way one needs to consider their place
in the right language game (cf. Ludwig Wittgenstein).

\section{Evaluation of Inferential Relations}

Models that learn inferential relations are the foundation for reasoning.
They represent a knowledge of the world. (The statements are descpritions
of pictures like in the SNLI corpus.) An important questions is, whether
the inferential relations are representing good inferences. How can
one contradictory statement be better than another one? Or how can
one entailment statement be better than another one? If we would take
this into account, we could introduce more fine grained logical connectors,
but then one might have to introduce an infinite number of logical
connectors. The approach that I propose is based on the web of inferential
relations that can be represented.

Meaning is not only represented in one inferential relation it is
part of a whole web of inferences. It is within a possible space of
reasoning and that includes more possible premises and consequences
of the statements. Evans et. al. (2018) introduce a model that learns
to recognize relations of possible ``entailment between logical formulas''.\cite{Evans2018}
It would be an interesting task to combine this model with the model
of Bowman et al. (2015a) that learns inferential relations of natural
language\cite{Bowman2015a} to widen the logical space of possible
relations. It could make explicit further inferences. This inferences
can lead to further statements, like mentioned above. It is correct
to state that a chair is not an animal, while it is incorrect to state
that a bird is not an animal. Of course, it relies also on the entailment
that birds are animals. Meaning can only be understood within a web
of inferences. This would allow the model to self-assess the goodness
of inferences and make it possible to discard bad inferences. 

Deductive reasoning fails to give an adequate account of (human) reasoning.
Inductive reasoning has strengths and with a lot of data and statistical
methods powerful tools have been developed, but does it give an account
of (human) reasoning and what could be considered as a good explanation
of something. To understand the goodness of inferences means to assess
the normative part of reasoning. 

\part*{References}

\bibliographystyle{IEEEtran}
\bibliography{BibliographyLogicalUnderstanding}

\end{document}